# Structural and mechanical properties of nitrogen-deficient cubic Cr-Mo-N and Cr-W-N systems

Liangcai Zhou[1], Fedor Klimashin[1], David Holec[2], and Paul H. Mayrhofer[1]

[1] Institute of Materials Science and Technology, Vienna University of Technology, A-1040 Vienna, Austria

[2] Department of Physical Metallurgy and Materials Testing, Montanuniversität Leoben, A-8700 Leoben, Austria

Abstract:

The tendency for nitrogen deficiency in cubic Cr-Mo-N and Cr-W-N solid solutions is predicted by a comprehensive evaluation of the lattice spacing, mixing thermodynamics, and elastic properties using first-principles calculations and experimentally confirmed by means of X-ray diffraction. A major conclusion is that these systems exhibit significant amount of N vacancies whose amount scales linearly with the TM content, hence making the $Cr_{1-x}TM_xN_{1-0.5x}$ chemical formula more precise and informative to describe the chemical composition of cubic Cr-Mo-N and Cr-W-N solid solutions as compared with the conventionally used $Cr_{1-x}TM_xN$. The cubic $Cr_{1-x}Mo_xN_{1-0.5x}$ and $Cr_{1-x}W_xN_{1-0.5x}$ solid solutions exhibit large positive mixing enthalpies towards isostructural phase decomposition into cubic B1-CrN and $\gamma$-$Mo_2N$ or $\gamma$-$W_2N$, respectively. Their ductility increases with increasing Mo or W content and both systems exhibit significantly direction-dependent Young's moduli over the entire composition range, even when using the approach to study their polycrystalline behavior. The excellent agreement between experimentally obtained lattice parameters, Mo- and W-dependent nitrogen content, elastic properties and their calculated values for our model descriptions, $Cr_{1-x}Mo_xN_{1-0.5x}$ and $Cr_{1-x}W_xN_{1-0.5x}$, allows to understand these complex material systems. Based on our results, we can conclude that their content of nitrogen vacancies scales with half of the alloying content Mo or W.

# 1. Introduction

The paramagnetic chromium nitride (CrN) with cubic B1 structure (NaCl prototype, space group $Fm\bar{3}m$) is one of the most important transitional metal nitrides (TMN), and not just used as a protective coating due to its superior wear, corrosion and oxidation resistance [1]. In order to further tune the structural, mechanical, and tribological properties to diverse service conditions, alloying with other elements has proven to be an effective concept [2-4]. Specifically, alloying molybdenum (Mo) or tungsten (W) into CrN coatings has been confirmed to be effective to improve the tribological properties and toughness [5, 6]. The ability of Mo and W to form Magnéli phase oxides can be used to prepare materials with solid lubricant potential, which significantly reduce the coefficient of friction for example [7, 8].

The present work is also motivated by the controversial results in literature. Recently, Quintela *et al.* [9] pointed out that even very small additions of only ~1 at.% of Mo or W to CrN, lead to a spontaneous phase segregation into Mo- and W-rich regions. Contrary, Kwang *et al.* [10] stated that cubic Cr-Mo-N coatings with Mo content less than 30.4 at.% are a substitutional solid solution of (Cr, Mo)N. Other reports state that deposited cubic Cr-Mo-N and Cr-W-N coatings undergo decomposition into cubic B1-CrN and γ-$Mo_2N$ or γ-$W_2N$ [5, 11-13].

The controversy is mainly due to the similarity of crystallographic structures of cubic B1-CrN and γ-$TM_2N$, which makes their XRD peaks very close and finally difficult to distinguish whether the coating is a mixture of B1-CrN+γ-$TM_2N$ or a cubic (Cr, TM)N solid solution, with TM being Mo or W [5]. Although the binary nitrides cubic B1-MoN and B1-WN are mechanically unstable [14], they are used as the boundary systems for cubic B1-(Cr, TM)N with nitrogen contents of constantly 50 at.%, as always reported [10, 15, 16]. By adjusting the $N_2$ gas partial pressure during the sputtering deposition process, the atomic ratio N/metal can be even above 1 for cubic (Cr, Mo)N or (Cr, W)N thin films [11, 17]. Employing a pair distribution function analysis of nearest neighbor distances between atoms, Shen *et al*. [18], stated that the excessive nitrogen atoms in overstoichiometric $WN_x$ thin films were present in the grain boundaries of $\gamma$-$W_2N$, not at the vacant octahedral sites.

The significant deviation for the lattice parameters of cubic $Cr_{1-x}TM_xN$ (TM = Mo or W) between theoretical and experimental data in our previous work [6] also suggests that the vacancies in these cubic $Cr_{1-x}TM_xN$ solid solutions play an indispensable and significant role. In order to clarify these controversies, we have performed a comprehensive investigation on understoichiometric cubic $Cr_{1-x}Mo_xN_y$ and $Cr_{1-x}W_xN_y$ solid solutions by the evaluation of the structural and mechanical properties using first-principles calculations. The results are compared with experimentally obtained lattice parameters of sputter deposited $Cr_{1-x}Mo_xN_y$ coatings.

## 2. Methods

### 2.1 Calculation details

Density Functional Theory (DFT) based calculations are performed using the Vienna Ab initio Simulation Package (VASP) [19, 20]. The ion-electron interactions are described by the projector augmented wave method (PAW) [21] and the generalized gradient approximation (GGA) as parameterized by Perdew–Burke–Ernzerhof (PBE) [22] is employed for the exchange-correlation effects.

In order to simulate the chemical disorder between Cr and TM atoms on the metal sublattice of the cubic B1 structure, and also the paramagnetic state induced by Cr atoms, we used the Special Quasi-random Structures (SQS) approach [23] as implemented and tested in our previous studied CrN-based systems [6, 24, 25]. The content of nitrogen vacancies is assumed to be dependent on the composition of γ-TM$_2$N, and the chemical formula Cr$_{1-x}$TM$_x$N$_{1-0.5x}$ (a linear combination of cubic B1-CrN and γ-TM$_2$N) is used to describe the cubic (Cr, TM)N solid solutions. The full compositional profile for the nitrogen vacancies, i.e., varying Cr/TM ratio on the metal sublattice and vacancy content on the nitrogen sublattice (which is similar to Refs. [26, 27]), was not accounted for. This is due to the fact that it is very difficult to incorporate more nitrogen atoms into vacant octahedral sites in γ-TM$_2$N [18, 28]. 2×2×2 supercells containing 64 atoms are used for the Cr$_{1-x}$TM$_x$N$_{1-0.5x}$ in the present work. The short range order parameters (SROs) are optimized for pairs at least up to the fifth coordination shell. The stress-strain method used in our previous work [6, 25] is employed to evaluate the elastic properties of Cr$_{1-x}$TM$_x$N$_{1-0.5x}$. Monkhorst-Pack grids [29] of 5×5×5 $k$-point mesh is used for the 64-atoms supercells. All the calculations were performed with plane wave cutoff energy of 500 eV, hence guaranteeing with an accuracy in the order of meV per atom.

## 2.2. Experimental details

All coatings in this study were deposited using a modified DC magnetron sputtering system Leybold Heraeus Z400. In order to vary the chemical composition, cubes of Cr (99.99% purity, 3x3x3mm) were uniformly arranged on the race track of a Mo target (99.99% purity and 75 mm in diameter), and reactively sputtered in a mixed N$_2$/Ar atmosphere onto single crystal silicon (100) and austenite substrates. Before the deposition, the chamber was evacuated to a high vacuum of 5·10$^{-4}$ Pa, and a constant total pressure of 0.35 Pa was kept during all sputtering processes. The substrate temperature of 450±20°C guaranteed for a dense columnar growth microstructure. This temperature regime corresponds to an homologous temperature T/T$_m$ ~0.3 [30] and thus belongs to the transition zone T according to Thornton's zone model [31]. The glow discharge current was set to 0.4 A.

Phase analyses were performed using an X-ray diffractometer Philips X'Pert with monochromized Cu $K_\alpha$ radiation. The lattice constants were obtained by fitting the peak positions in the diffraction patterns, which are recorded in Bragg-Brentano geometry. Elemental composition was determined by means of energy dispersive X-ray spectroscopy (EDXS) using a FEI Quanta 200 FEGSEM scanning electron microscope equipped with a Schottky emitter, which provides an excellent spatial resolution of about 2 nm. Indentation moduli were measured with a UMIS unit using a Berkovich diamond tip and applying loads between 3 and 45 mN. The Young's moduli were evaluated according to the procedure after Oliver and Pharr [32], and plotted as a function of the penetration depth. The best estimation of the film-only elastic modulus can be achieved by extrapolating the measured data back to a zero penetration depth [33]. In order to verify the obtained results of the elastic moduli, series of measurements were carried out on different reference samples with well-known values of Young's modulus. The coatings on austenite substrates were used due to the better adhesion, since even microscopic delamination will influence the evaluation of the Young's modulus.

## 3. Results and discussion

Figure 1 presents the crystal structures of cubic B1-CrN, γ-TM$_2$N (TM = Mo and W), β-TM$_2$N, and t-γ-TM$_2$N. γ-TM$_2$N has a defective cubic B1 structure with 50% disordered vacancies in the nitrogen sublattice [34]. β-TM$_2$N (space group $I4_1/amd$) is a low-temperature tetragonal phase with an ordering of vacancies on the non-metal sublattice from γ-TM$_2$N [35]. Tetragonal t-Mo$_2$N is constructed from perfect cubic B1-TMN by removing 50% of the nitrogen atoms in the [100] and [010] directions in order to obtain 50% vacancies on the non-metal sublattice [36].

Three different structural configurations are tested for γ-TM$_2$N and the results are listed in table 1, where γ-TM$_{32}$N$_{16}$ denotes the SQS configuration. It clearly shows that different structural configurations only little influence the total energies, lattice spacing, and bulk moduli, therefore they are not further considered. Especially, the experimental bulk moduli and the lattice parameters of γ-TM$_2$N [18, 37] show a good agreement with calculated results.

We present the nitrogen content for Cr-Mo-N coatings as a function of their Mo content, see Fig. 2a, and their lattice parameters, Fig. 2b. Figure 2a clearly shows that the nitrogen content within our Cr-Mo-N coatings strongly depends on their Mo content. The experimental observations (solid symbols) can nicely be described with our calculated model solid solutions (open symbols). Based on this agreement we can conclude that the nitrogen vacancies have to be taken into account for these material systems. The calculated lattice parameters of stoichiometric cubic Cr$_{1-x}$TM$_x$N (TM = Mo and W) show a significant deviation from experimental data [4, 11, 12, 16, 34] and the deviation increase with increasing Mo or W content [6], see Fig. 2b. On the other hand, the calculated results for cubic Cr$_{1-x}$TM$_x$N$_{1-0.5x}$ with N vacancies yield a good agreement with the experimental results. The lattice parameters of cubic Cr$_{1-x}$Mo$_x$N$_{1-0.5x}$ and Cr$_{1-x}$W$_x$N$_{1-0.5x}$ are close to each other, and show a linear Vegard-like behavior [38]. This indicates that the nitrogen content in the cubic Cr-Mo-N and Cr-W-N solid solutions is linearly dependent on the fraction of γ-TM$_2$N and hence varies along the CrN+γ-TM$_2$N tie line, which has recently been confirmed by an independently experimental work [39], which is not published yet. Therefore, we propose using chemical formula Cr$_{1-x}$TM$_x$N$_{1-0.5x}$ to describe the chemical composition of cubic Cr-Mo-N and Cr-W-N solid solutions. Here we should notice that under some extreme conditions, it is still possible to prepare Cr$_{1-x}$TM$_x$N materials, but just with a limited maximum Mo or W content, for example, using a very high (pure) N$_2$ pressure to prepare the these coatings. While the theoretical and experimental lattice constants yield an excellent agreement for $Cr_{1-x}Mo_xN_{1-0.5x}$, the experimental lattice constants of $Cr_{1-x}W_xN_{1-0.5x}$ show a small positive deviation from the theoretical results, which is more likely due to the excessive nitrogen atoms at the grain boundaries for overstoichiometric $WN_x$ thin films [18]. As mentioned above, Cr-Mo-N and Cr-W-N solid solutions exhibit a strong tendency for N substoichiometry, and hence using $Cr_{1-x}TM_xN_{1-0.5x}$ as the chemical formula for cubic Cr-Mo-N and Cr-W-N solid solutions is more precise than the conventionally used $Cr_{1-x}TM_xN$ as it expresses the correct information about the chemical composition of cubic Cr-Mo-N and Cr-W-N solid solutions.

Previous theoretical calculations have illustrated that the mixing enthalpies of $Cr_{1-x}TM_xN$ with respect to cubic B1-CrN and TMN are in contrast with experimental observations [6], and thus cannot be used to evaluate the phase stability of cubic Cr-Mo-N and Cr-W-N systems. In an earlier work, we have argued that this is because cubic B1-MoN and B1-WN are mechanically unstable, and therefore lead to the anomalous negative mixing enthalpies [6]. Therefore, we have suggested that using $CrN+TM_2N+N_2$ as the reference state for evaluating the mixing enthalpies of $Cr_{1-x}TM_xN$ is more meaningful [6]. Although the mixing enthalpies of $Cr_{1-x}TM_xN$, when using $CrN+TM_2N+N_2$ as the reference state, predict the possibility of isostructural decomposition, the description is not really correct, as the large substoichiometry of nitrogen in Cr-Mo-N and Cr-W-N solid solutions is not taken into account.

Therefore, in addition to our previous studies also the mixing enthalpies of cubic $Cr_{1-x}Mo_xN_{1-0.5x}$ and $Cr_{1-x}W_xN_{1-0.5x}$ are evaluated with respect to cubic B1-CrN and $\gamma$-$TM_2N$ (Fig. 3). Consequently, these materials correspond to the quasi-binary tie line $CrN$–$Mo_2N$ and $CrN$–$W_2N$. Inspired by recent works [24, 40], showing that the strong electron correlation effects play a crucial role on the phase stability, the local density approximation (LDA) plus a Hubbard U-term method within the framework of the Dudarev formulation [41, 42] is employed. This allows to consider the impact of strong electron correlation effects on the mixing enthalpy, together with 4×4×4 supercells containing 128 atoms ($k$-point meshes: 3×3×3). The black lines denote the mixing enthalpies of cubic $Cr_{1-x}TM_xN_{1-0.5x}$ from GGA scheme using 64-atoms supercells. Their mixing enthalpies evaluated from LDA+U scheme, with the value of effective U-J = 3eV using 128-atoms supercells, are represented by dotted lines.

The cubic $Cr_{1-x}Mo_xN_{1-0.5x}$ and $Cr_{1-x}W_xN_{1-0.5x}$ systems exhibit a similar mixing behavior, with the same maximum mixing enthalpies of about 0.11 and 0.17 eV/atom from GGA and LDA+U schemes, respectively. The higher mixing enthalpies for cubic $Cr_{1-x}Mo_xN_{1-0.5x}$ and $Cr_{1-x}W_xN_{1-0.5x}$ systems from LDA+U scheme originate from the impact of considering strong electron correlations effects, which make the $d$-states more localized and finally weaken the hybridization between the neighboring metal-metal bonds as compared to GGA exchange-correlation approximation. The large positive mixing enthalpies of cubic $Cr_{1-x}Mo_xN_{1-0.5x}$ and $Cr_{1-x}W_xN_{1-0.5x}$ are mainly due to the significantly larger covalent radii of Mo (1.45 Å) and W (1.46 Å) as compared with Cr (1.27 Å) [43]. Although our data explicitly address only certain content of nitrogen vacancies (corresponding to the $Cr_{1-x}TM_xN_{1-0.5x}$ chemical formula), they represent the lower limit of the mixing enthalpies of cubic $Cr_{1-x}Mo_xN_{1-y}$ (with $y < 0.5x$). This conclusion is also based on experimental observations showing that it is extremely difficult to incorporate more nitrogen atoms into γ-$TM_2N$. Actually, for nitrogen contents above 50 at.%, the γ-$TM_2N$ transforms into other structures [28, 37]. The large positive mixing enthalpies for $Cr_{1-x}Mo_xN_{1-0.5x}$ and $Cr_{1-x}W_xN_{1-0.5x}$ are comparable to the well-investigated $Ti_{1-x}Al_xN$ system with a maximum mixing enthalpy of about 0.11 eV/atom, which experiences a spontaneous spinodal decomposition into coherent cubic domains of AlN and Ti-enrich $Ti_{1-x}Al_xN$ [44]. This implies that cubic Cr-Mo-N and Cr-W-N systems show great tendency towards isostructural phase decomposition and by near-equilibrium deposition techniques it will be difficult to synthesize single phase cubic $Cr_{1-x}Mo_xN_{1-0.5x}$ and $Cr_{1-x}W_xN_{1-0.5x}$ solid solutions. This is consistent with recent experimental observations, showing that even for the lowest concentrations (~1 at.%), spontaneous phase segregation into Mo- and W-rich regions occurs [9]. We envision that the deposited Cr-TM-N thin films were misleadingly evaluated as single phase cubic solid solutions [10-12], because it is difficult to distinguish the cubic B1-CrN and γ-$TM_2N$ just from conventional XRD measurements due to their similarity in lattice parameters. Based on our results we can conclude, that the huge positive mixing enthalpies clearly provide a strong explanation for the large tendency for isostructural phase decomposition in Cr-Mo-N and Cr-W-N systems. The B1-CrN and γ-$Mo_2N$ or γ-$W_2N$ phases may not be mixed with each other under near-equilibrium conditions.

However, by physical vapor deposition, especially when using only moderate substrate temperatures (below 0.3 of the melting point of the prepared coating), solid solutions far from the thermodynamic equilibrium (i.e., metastable and even thermodynamically unstable) can be realized [3]. The calculated elastic constants $C_{11}$, $C_{12}$, and $C_{44}$, for cubic $Cr_{1-x}Mo_xN_{1-0.5x}$ and $Cr_{1-x}W_xN_{1-0.5x}$ solid solutions (Fig. 4), fulfill the corresponding Huang-Born stability criteria [45]. Consequently, cubic $Cr_{1-x}Mo_xN_{1-0.5x}$ and $Cr_{1-x}W_xN_{1-0.5x}$ are mechanically stable over the entire composition range. This is a further strong argument (in addition to the excellent agreement between calculated and measured nitrogen content and the calculated mixing enthalpy), that for these material systems the content of nitrogen vacancies scales with 0.5 of the Mo or W content, hence their chemical formula is $Cr_{1-x}TM_xN_{1-0.5x}$. The calculated elastic constant $C_{12}$ increases with the content of γ-$TM_2N$, while $C_{44}$ keeps an almost constant value in the whole composition range. Consequently, the Cauchy pressure, $C_{12}$-$C_{44}$, increases with increasing Mo or W content and thereby, also the metallic bonding character and ductility of $Cr_{1-x}TM_xN_{1-0.5x}$.

Deposited thin films usually have a fibrous growth texture with a preferred orientation along a certain <$hkl$> direction perpendicular to the substrate surface [46]. Therefore, we calculated directionally dependent Young's moduli, $E_{hkl}$, for cubic $Cr_{1-x}Mo_xN_{1-0.5x}$ and $Cr_{1-x}W_xN_{1-0.5x}$ compositions in <100>, <110>, and <111> directions, Fig. 5. Both cubic solid solutions, $Cr_{1-x}Mo_xN_{1-0.5x}$ and $Cr_{1-x}W_xN_{1-0.5x}$, exhibit direction dependent Young's moduli in the whole composition range, with <100> direction being significantly stiffer than the other directions. This behavior follows the single crystal elastic constant $C_{ij}$, Fig. 4, due to the strongest contribution of $C_{11}$ to $E_{100}$. The calculated $E_{100}$ data are in excellent agreement to the experimentally obtained Young's moduli of our coatings with a strong <100> growth texture. The literature data, with a reported preferred <111> orientation [12, 13], are in excellent agreement with our $E_{111}$ calculations. Our data clearly shows that the Young's moduli can be significantly different for various directions. Consequently, this needs to be considered when comparing experimental and theoretical data [47].

## 4. Conclusions

The nitrogen substoichiometry in cubic Cr-Mo-N and Cr-W-N solid solutions was systemically studied by first-principles calculations and compared with experimental data of the lattice spacing, mixing thermodynamics, and elastic properties. Cubic Cr-Mo-N and Cr-W-N systems intrinsically contain a significant amount of vacancies on N sublattice. The content of vacancies increases with increasing Mo or W fraction. Based on our comprehensive investigations we propose a new chemical formula for cubic Cr-Mo-N and Cr-W-N systems, $Cr_{1-x}TM_xN_{1-0.5x}$, which reflects the increasing amount of vacancies on the N sublattice, and consequently is more precise than with the conventionally used $Cr_{1-x}TM_xN$. The large positive mixing enthalpies of cubic $Cr_{1-x}Mo_xN_{1-0.5x}$ and $Cr_{1-x}W_xN_{1-0.5x}$ indicate a significant thermodynamic driving force for isostructural phase decomposition into cubic B1-CrN and γ-$TM_2N$. The elastic constants show that cubic $Cr_{1-x}Mo_xN_{1-0.5x}$ and $Cr_{1-x}W_xN_{1-0.5x}$ solid solutions are mechanically stable in the whole composition range and the positive Cauchy pressure, $C_{12}-C_{44}$, indicates the increasing ductility with increasing content of Mo or W. An excellent agreement of Young's moduli is obtained between theoretical and experimental results once the coating's texture is considered.

**Acknowledgements**

The financial support by the START Program (Y371) of the Austrian Science Fund (FWF) is gratefully acknowledged. The computational results presented have been achieved using the Vienna Scientific Cluster (VSC).

Table 1. Calculated total energy, $E_{total}$, lattice parameter, $a$, volume per atom, and bulk modulus, $B$. The experimental data is presented for comparison.

| | γ-Mo$_{32}$N$_{16}$ | β-Mo$_2$N | t-Mo$_2$N | γ-Mo2N | γ-W$_{32}$N$_{16}$ | β-W$_2$N | t-W$_2$N | γ-W2N |
|---|---|---|---|---|---|---|---|---|
| $E_{total}$ (eV/atom) | -10.213 | -10.248 | -10.191 | | -11.472 | -11.481 | -11.406 | |
| Lattice parameter (Å) | $a$=4.19 | $a$=4.27 $c$=8.04 | $a$=4.22 $c$=4.18 | 4.16 [34] | $a$=4.20 | $a$=4.30 $c$=7.98 | $a$=4.22 $c$=4.19 | 4.19 [18] |
| $V$ (Å$^3$/atom) | 12.26 | 12.26 | 12.44 | 12.00 [34] | 12.35 | 12.35 | 12.53 | 12.26 [18] |
| $B$ (GPa) | 298 | 288 | 303 | 301±7 [37] | 346 | 340 | 340 | |

Figure captions

Fig. 1. (Color online) Crystallographic structures of cubic B1-CrN, γ-TM$_2$N, β-TM$_2$N, and t-TM$_2$N.

Fig. 2. (Color online) (a) The nitrogen content of deposited Cr-Mo-N thin films (solid symbols) as a function of the Mo content. The nitrogen compositions used in the calculations (open symbols) are added for comparison. (b) The lattice parameters of Cr$_{1-x}$TM$_x$N (half-open symbols) and Cr$_{1-x}$TM$_x$N$_{1-0.5x}$ (solid symbols) (TM = Mo and W) as function of TM fraction on metal sublattice, together with experimental values (open symbols) [4, 11, 12, 16, 34] for comparison.

Fig. 3. (Color online) Calculated isostructural mixing enthalpies of Cr$_{1-x}$Mo$_x$N$_{1-0.5x}$ (open circle symbols with red lines) and Cr$_{1-x}$W$_x$N$_{1-0.5x}$ (solid square symbols with black lines) as functions of the TM$_2$N content. The solid lines denote the results from the GGA scheme with 64-atoms supercells, while the dotted lines denote the results derived from the LDA+$U$ ($U$-$J$=3 eV) scheme with 128-atoms supercells.

Fig. 4. (Color online) Calculated single crystal elastic constants $C_{11}$, $C_{12}$, and $C_{44}$ of Cr$_{1-x}$Mo$_x$N$_{1-0.5x}$ and Cr$_{1-x}$W$_x$N$_{1-0.5x}$.

Fig. 5. Young's moduli in the <100> (square symbols), <110> (triangle symbols), and <111> (circle symbols) directions of Cr$_{1-x}$TM$_x$N$_{1-0.5x}$ (TM = Mo and W) as functions of the TM$_2$N content. The experimental Young's moduli (solid symbols) [12, 13] are presented for comparison.

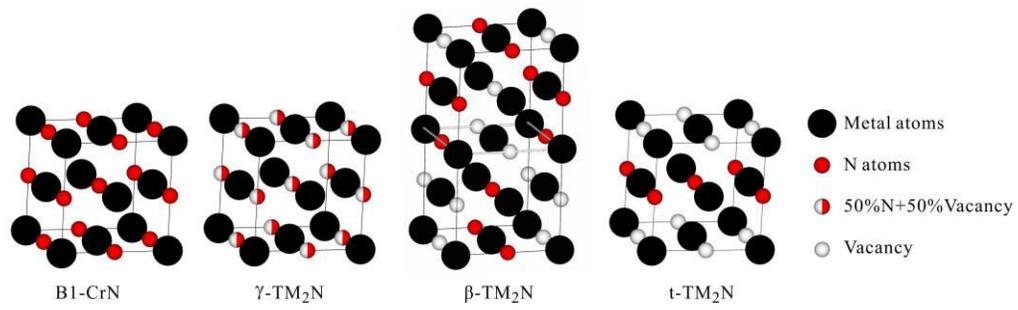

Fig. 1.

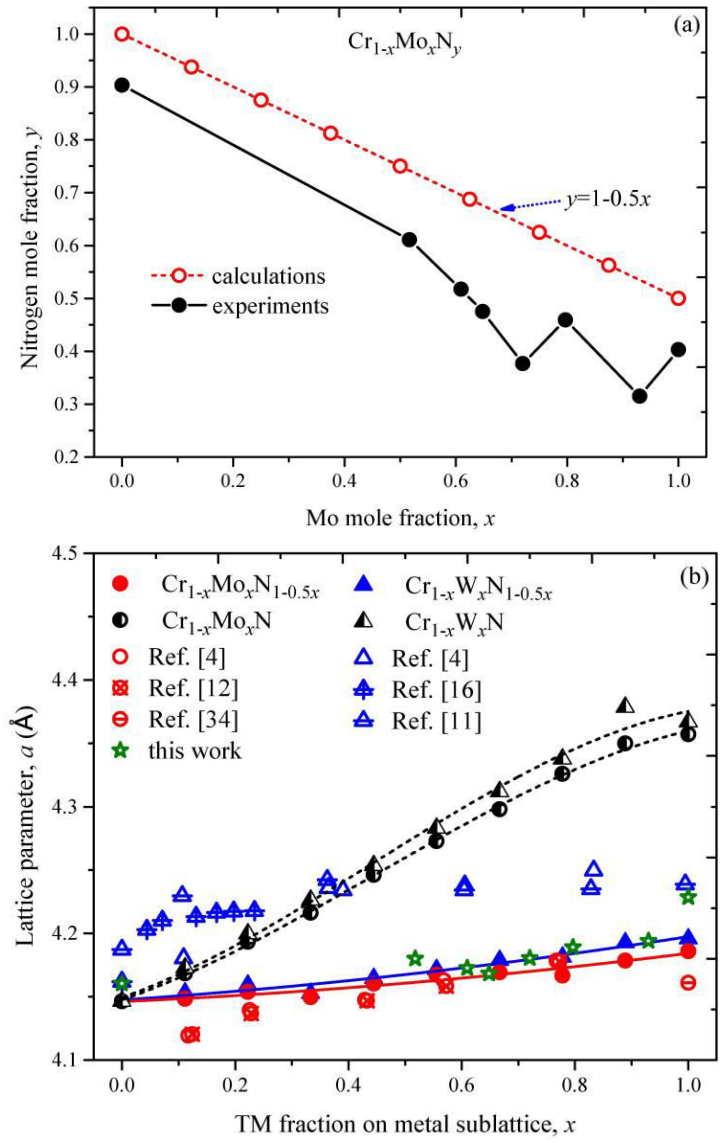

Fig. 2.

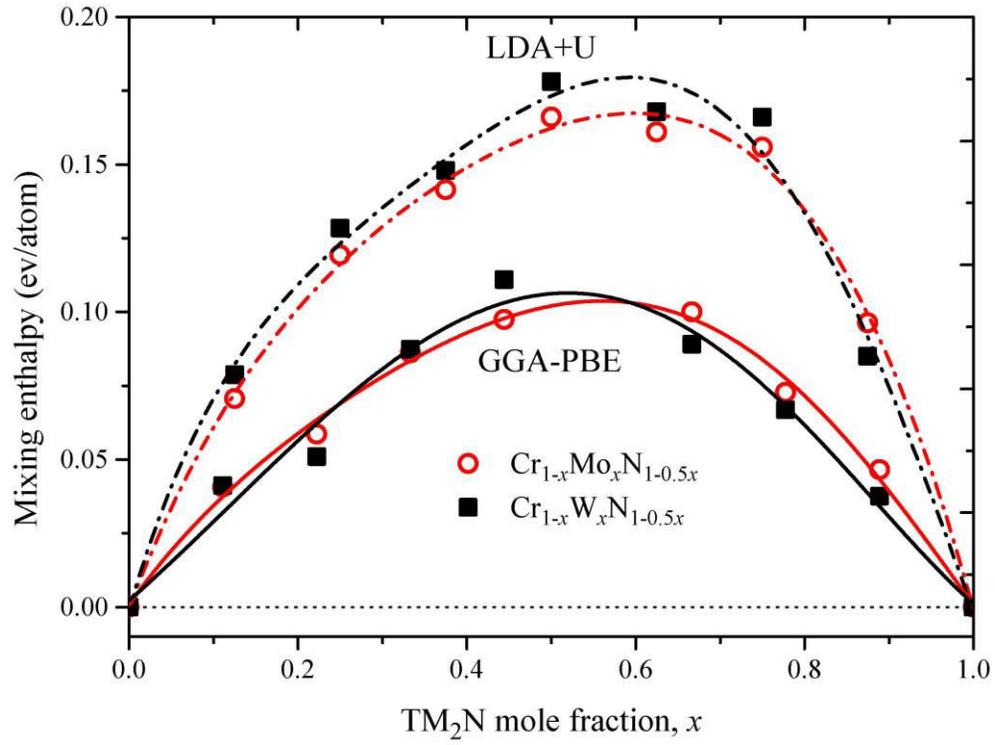

Fig. 3.

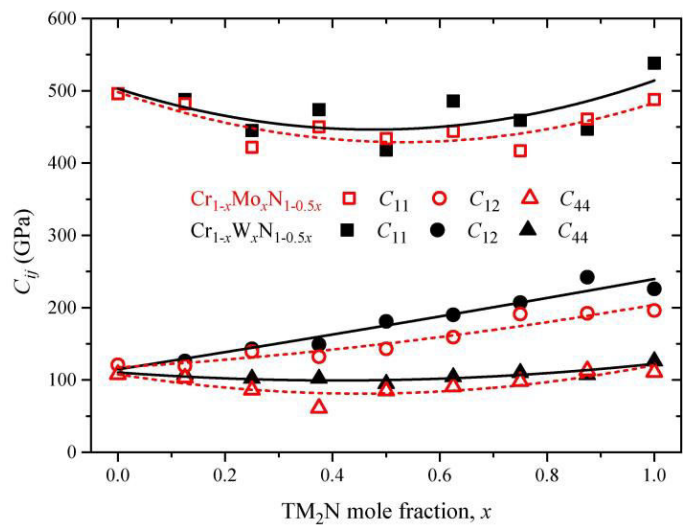

Fig. 4

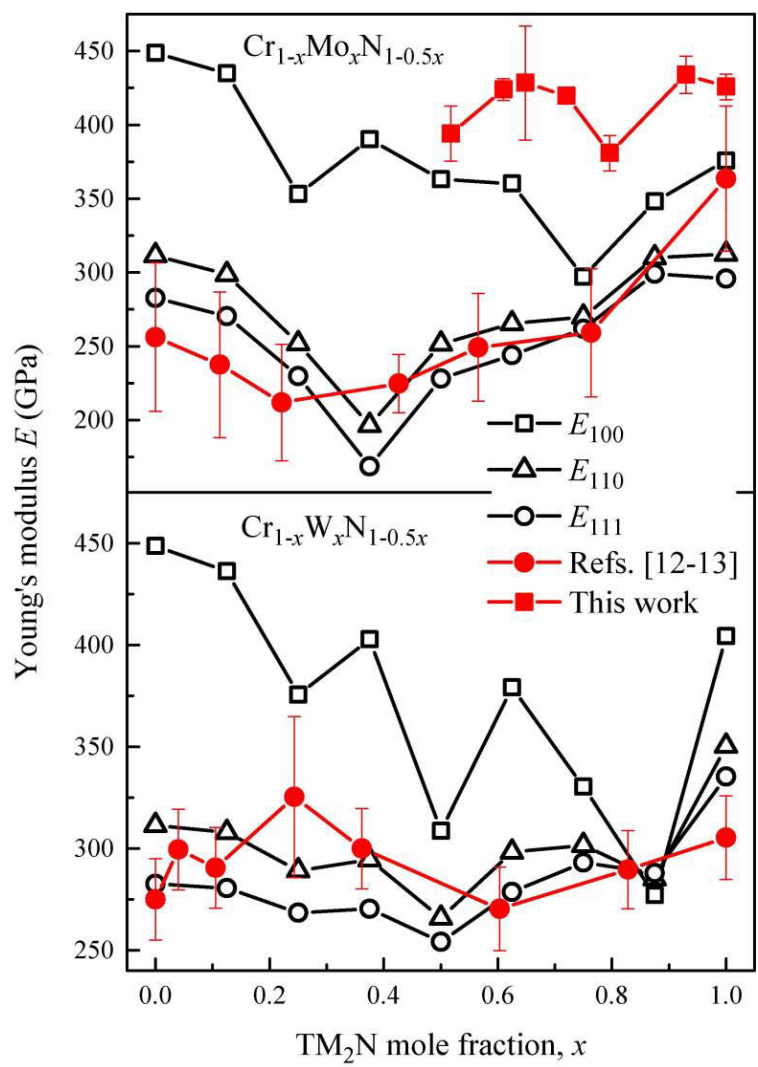

Fig. 5.